\documentclass[12pt,a4paper]{article}

\usepackage{amsmath}
\usepackage{amssymb}
\usepackage{nicefrac}
\usepackage{mathrsfs}
\usepackage{braket}
\usepackage{physics}
\usepackage[version=4]{mhchem}

\usepackage{graphicx}
\usepackage{float}
\usepackage{dcolumn} 
\newcolumntype{.}{D{.}{.}{1}}
\usepackage{booktabs,threeparttable}

\usepackage{microtype}
\usepackage{lmodern}

\usepackage{sectsty}
\sectionfont{\sffamily}
\subsectionfont{\sffamily}
\subsubsectionfont{\sffamily}
\usepackage{dsfont}

\usepackage[
hidelinks,
colorlinks=false,
breaklinks=true,
linkcolor=black,
menucolor=black,
filecolor=black,
urlcolor =black,
citecolor=black
]{hyperref}

\usepackage{color}
\usepackage{cite}
\usepackage{lineno} 
\usepackage{tikz,pgfplots}

\usepackage{xcolor}
\usepackage[normalem]{ulem}

\begin{document}
\begin{center}
\vspace*{0.5cm}

{\Large\bf
Multi-State Formulation of the Frozen-Density Embedding Quasi-Diabatization Approach
}

\vspace{2cm}

{\large Patrick Eschenbach$^{\parallel}$, Denis G. Artiukhin$^{\dagger}$\footnote{Email: artiukhin@chem.au.dk},
and Johannes Neugebauer$^{\parallel}$\footnote{Email: j.neugebauer@uni-muenster.de} \\[2ex]
}

\vspace{0.5cm}

$^{\parallel}$
Theoretische Organische Chemie, Organisch-Chemisches Institut and
Center for Multiscale Theory and Simulation,
Westf\"alische Wilhelms-Universit\"at M\"unster, Corrensstra{\ss}e 40,
48149 M\"unster, Germany \\[1ex]

$^{\dagger}$
Department of Chemistry, Aarhus Universitet, DK-8000 Aarhus, Denmark \\[1ex]

\vspace{0.5cm}
\end{center}

\vfill

\begin{tabbing}
Date:   \quad\= \today \\
\end{tabbing}

\newpage

\begin{abstract}
We present a multi-state implementation of the recently developed FDE-diab methodology [\textit{J.\ Chem.\
Phys.}, 148 (\textbf{2018}), 214104] in the \textsc{Serenity} program. 
The new framework extends the original approach such that any number of charge-localized quasi-diabatic states can be coupled, giving an access to calculations of ground and excited state spin-density distributions as well as to excitation energies.
We show that it is possible to obtain results similar to those from
correlated wave function approaches such as the  complete active space self-consistent field method at much lower computational effort.
Additionally, we present a series of approximate computational schemes, which further
decrease the overall computational cost and systematically converge to the full FDE-diab solution.
The proposed methodology enables computational studies on spin-density distributions and related properties for large molecular systems of biochemical interest.

\end{abstract}

\newpage

\section{Introduction} \label{sec:intro}

The spin density is commonly defined as the difference of $\alpha$ and $\beta$ electron densities 
and, thus, represents the excess of $\alpha$ electrons in a radical system. 
The accurate determination of the spin density is of high interest for both theory and experiment.
Thus, in spin density functional theory (DFT), which is frequently used for open-shell molecular systems, 
both the electron and spin density are 
required for obtaining the exact ground-state energy~\cite{jacob2012}.
Negative areas of the spin density are also an important
ingredient in calculations of 
total squared spin operator expectation values
$\expval{S^2}$ (within the exchange local spin 
density approximation, see Ref.~\cite{wang1995}), 
which can subsequently be used for the determination of magnetic exchange coupling constants~\cite{yamaguchi1986,soda2000}.
The latter play a major role in predicting
magnetic interactions
between, e.g., building blocks of
organic radical crystals~\cite{deumal2002,dresselhaus2016}. 
Experimentally, spin-density maps can directly be measured with polarized neutron 
diffraction~\cite{brown1989,papoular1990,zheludev1994}. 
An indirect access to spin densities can be obtained with electron paramagnetic resonance (EPR) spectroscopy via
the isotropic part of EPR hyperfine coupling constants, which depends on the spin-density contributions at 
atomic nuclei~\cite{nmr_epr_book}.
Specialized methods such as the solid-state photochemically induced dynamic nuclear polarization nuclear magnetic resonance 
and electron nuclear double resonance spectroscopy are also widely used for the determination of spin-density maps 
(for applications to cofactors from photosynthetic reaction centers, see 
Refs.~\cite{dolphin1970,fajer1979,lend1993,kaess1995,lubitz2002,daviso2009,janssen2010,najdanova2015,janssen2018}).

Although many experimental techniques can be used for the accurate determination of spin densities, theoretical computations of 
these molecular properties are challenging for most electronic structure methods~\cite{bally1999,reiher2009,boguslawski2011}. 
Highly correlated \textit{ab initio} wave function approaches are often used to calculate accurate spin 
densities and related properties 
such as EPR hyperfine coupling constants~\cite{radon2010,kossmann2010,boguslawski2012,lan2014,shiozaki2016}. 
Unfortunately, these methods are computationally very demanding and, thus, 
not feasible for large molecular systems.
This limitation is often lifted by using Kohn--Sham density functional theory (KS-DFT)~\cite{jacob2012}.
However, it is well known that KS-DFT suffers from the self-interaction error~\cite{sanchez2006,grafenstein2004,cohen2007}
, which often leads to overdelocalized spin-density distributions, especially in cases of non-covalently bonded molecular systems~\cite{cohen2008,solo2012}.
This issue can be mitigated
employing hybrid exchange--correlation (XC) functionals, which incorporate
exact Hartree--Fock (HF) exchange~\cite{boguslawski2011}.
On the downside, however, increasingly large areas of negative spin density and increasing spin contamination are
observed in this case~\cite{herrmann2005,herrmann2006,cohen_tozer_2007,Radom2008,pavanello2011,solovyeva2012,arti2017}. 
Additionally, the use of hybrid XC functionals can lead to ``broken-symmetry-like''
solutions providing qualitatively wrong spin densities 
as was shown for iron-nitrosyl complexes in Ref.~\cite{boguslawski2011}.

The overdelocalization problem of KS-DFT can pragmatically be avoided 
by using the Frozen-Density Embedding (FDE) formalism~\cite{weso1993,weso1996,weso1997b,weso2006}.
FDE employs an additional non-additive kinetic energy potential (NAKP) and is, therefore, an approximation to KS-DFT.
When the exact NAKP is used within FDE, both FDE and KS-DFT provide 
the same charge- and spin-density distributions (given that the same XC functional and 
basis set are applied)~\cite{pavanello2011,solovyeva2012,schnieders2018}. 
In practice, however, approximate NAKPs are commonly employed. 
These approximate potentials are often repulsive in the intermolecular regime and could, therefore, prevent charge- and spin-density delocalization 
over subsystem boundaries~\cite{solo2012}.
Although the use of a supermolecular basis set increases the degree of delocalization, this effect is rather modest. 
In fact, it can even be argued that the use of FDE leads to the opposite problem of overly localized spin-density distributions~\cite{arti2018}.

Recently, the FDE-diab approach for accurate calculations of spin densities was proposed~\cite{arti2018}. 
This methodology is based on the previously known FDE-ET~\cite{pavanello2011,pavanello_voorhis2013,solo2014,ramos_papa_pavall2015}, 
and employs ideas of constructing charge- and spin-localized states
with FDE and coupling them in a configuration interaction (CI)-like manner.
It was demonstrated that this procedure avoids the consequences of KS-DFT spin overdelocalization error 
and does not lead to overly localized spin distributions characteristical for FDE. 
FDE-diab spin densities were compared against 
accurate \textit{ab initio} computations~\cite{arti2018} and 
experimental measurements~\cite{arti2020,arti2021}.
A good qualitative agreement was found in all cases proving that the proposed methodology 
is accurate and effective.
However, the program implementation available to date is limited to only two quasi-diabatic states. 
This severely restricts the applicability of FDE-diab and prevents one from calculations of supramolecular systems, 
where spin may naturally be delocalized over multiple fragments.
In this work, we aim at filling this gap and present a multi-state extension of FDE-diab. 
To that end, we describe the underlying theory of this new approach, thoroughly test it on a number of model molecular systems, 
and demonstrate approximate schemes which decrease the overall computational cost and extend the applicability of the method 
towards larger molecular systems. We show that the proposed methodology 
is not limited by calculations of ground electronic state spin-density distributions but
can also be used to obtain 
electronic couplings, excitation energies, as well as spin distributions in excited electronic states.

This work is structured as follows. 
In Sec.~\ref{sec:theory}, the underlying theory of multi-state FDE-diab and derived approximate approaches
is briefly summarized.
Computational details are given in
Sec.~\ref{sec:comput_details} and followed by results presented in Sec.~\ref{sec:results}. 
Finally, conclusions are drawn in Sec.~\ref{sec:conclusion}.

\section{Theory} \label{sec:theory}

In the following, the theory behind the FDE-diab 
approach~\cite{pavanello2011,pavanello_voorhis2013,solo2014,ramos_papa_pavall2015,arti2018} is
briefly introduced with a focus on the use of multiple quasi-diabatic states.
Special attention is given to various approximations which can be invoked within the FDE-diab framework.
For a more detailed explanation of underlying theoretical concepts, 
the reader is referred to the original works and Sec.~S1 of the Supporting Information (SI).

\subsection{FDE-diab for Multiple Electronic States}

Within the multi-state formulation of FDE-diab, each adiabatic electronic wave function $\Psi_m$ is represented
as a linear combination of $n$ quasi-diabatic states $\Phi_i$,
\begin{equation} \label{eq:linearcombination}
\Psi_m = B_{1m} \Phi_1 + B_{2m} \Phi_2 + \dots + B_{nm} \Phi_n,
\end{equation}
where $B_{im}$ are linear combination coefficients.
The quasi-diabatic states $\Phi_i$ are constructed as direct products
(similar to those described in Refs.~\cite{migliore2009,migliore2011}) of subsystem Slater determinants $\ket{A}$, $\ket{B}$, etc.,
\begin{equation} \label{eq:diabaticstate}
\Phi_i = \ket{A}\otimes\ket{B}\otimes\dots= \ket{AB\dots},
\end{equation}
which are obtained from prior FDE~\cite{weso1993} calculations of the corresponding subsystems $A$, $B$, etc. 
In these FDE calculations, an electronic charge can be localized at a particular molecular fragment giving rise to different
charge-localized states $\Phi_i$. Additionally, subsystems in their electronically excited states can be considered~\cite{solo2014}.
To obtain molecular orbitals (MOs) comprising the mentioned above Slater determinants, the so-called Kohn--Sham equations with constrained electron density (KSCED)~\cite{weso1993,weso1999,weso2006},
\begin{equation}
    \bigg[ -\frac{\nabla^2_i}{2} + \upsilon^{I,\sigma}_{\text{KS}}[\rho_I^{\sigma}](\vec{r}\,) +  \upsilon^{I,\sigma}_{\text{emb}}[\rho_I^{\sigma};\rho^{\text{tot},\sigma}](\vec{r}\,)
    \bigg]\phi_{i}^{I,\sigma}(\vec{r}\,) = \epsilon_{i}^{I,\sigma} \phi_{i}^{I,\sigma}(\vec{r}\,),
\label{eq:KSCED}
\end{equation}
need to be solved. Here, $-\frac{\nabla^2_i}{2}$ is the one-electron kinetic energy operator, 
$\upsilon^{I,\sigma}_{\text{KS}}[\rho_I^{\sigma}](\vec{r}\,)$ is the Kohn--Sham (KS) potential, 
and $\upsilon^{I,\sigma}_{\text{emb}}[\rho_I^{\sigma};\rho^{\text{tot},\sigma}](\vec{r}\,)$ is
the embedding potential. 
$\phi_{i}^{I,\sigma}(\vec{r}\,)$ and $\epsilon_{i}^{I,\sigma}$ are the KS-like MOs and orbital energies, respectively.
$I$ is the subsystem index, and $\sigma = \alpha$ or $\beta$ is a spin label.

With the quasi-diabatic states $\Phi_i$ constructed, the linear combination coefficients $B_{im}$ from Eq.~(\ref{eq:linearcombination}) have to be found
to calculate the adiabatic wave functions $\Psi_m$. To that end, a generalized eigenvalue problem in the basis of the quasi-diabatic states $\Phi_i$ 
is solved,
\begin{equation} \label{eq:eigenvalueproblem}
\mathbf{HB=SBE},
\end{equation}
where $\mathbf{H}$ and $ \mathbf{S}$ are the $n \times n$ Hamilton and overlap matrices, respectively. 
The matrix $\mathbf{B}$ contains the linear combination coefficients $B_{im}$, and $\mathbf{E}$ is a diagonal matrix of electronic energies $E_m$.
The elements $S_{ij}$ of the overlap matrix $\mathbf{S}$ are defined as~\cite{loewdin1955,mayer2003},
\begin{equation}  \label{eq:overlapelement}
    S_{ij} = \braket{\Phi_i}{\Phi_j} = \det\left(\mathbf{S}^{(ij), \alpha}\right) \det\left(\mathbf{S}^{(ij), \beta}\right).
\end{equation}
Here, $\mathbf{S}^{(ij), \sigma}$ are the MO overlap matrices with elements $\left(\textbf{S}^{(ij), \sigma}\right)_{kl}$ given by
\begin{equation} \label{eq:motransoverlap}
    \left(\textbf{S}^{(ij), \sigma}\right)_{kl} = \Braket{\phi_{k}^{(i),\sigma}|\phi_{l}^{(j),\sigma}}.
\end{equation}
The orbitals $\phi_{k}^{(i),\sigma}$ and $\phi_{l}^{(j),\sigma}$ belong to the quasi-diabatic states $\Phi_i$ and $\Phi_j$, respectively.
Because a monomer basis set is used for construction of quasi-diabatic states in FDE-diab (for examples, see Refs.~\cite{pavanello2011,arti2018}), 
MOs of different subsystems are not orthogonal to each other. Therefore, the off-diagonal matrix elements from Eq.~(\ref{eq:motransoverlap}) differ from zero.
The Hamilton matrix elements $H_{ij}$ from Eq.~(\ref{eq:eigenvalueproblem}) are approximated by calculating an KS energy functional of the transition electron density 
$\rho^{(ij)}(\vec{r}\,)$ and scaling it by the overlap $S_{ij}$~\cite{pavanello2011,pavanello_voorhis2013,solo2014,ramos_papa_pavall2015},
\begin{equation}
    H_{ij} = \mel**{\Phi_i}{\hat{H}}{\Phi_j} \approx E\left[\rho^{(ij)}(\vec{r}\,)\right] S_{ij}. \label{eq:hamiltonelement}
\end{equation}
The $\alpha$- and $\beta$-components of the transition electron density $\rho^{(ij)}(\vec{r}\,)$ are obtained by applying
the corresponding electron density operators to quasi-diabatic wave functions $\Phi_i$ such that
\begin{equation}
	\rho^{(ij),\sigma}(\vec{r}\,) = \frac{\bra{\Phi_i}  \hat{\rho}^{\sigma} \ket{ \Phi_j}}{\braket{\Phi_i}{\Phi_j}} \label{eq:trans-dens1},
\end{equation}
where expressions for operators $\hat{\rho}^{\sigma}$ are given by
\begin{equation} \label{eq:sigma_dens_oper}
    \hat{\rho}^{\sigma}  = \sum_{l=1}^{N_{\mathrm{el}}} \left[ \frac{1}{2} \hat{\mathds{1}}_l \pm \hat{s}_{z,l} \right]\delta(\vec{r}_l - \vec{r}\,).
\end{equation}
Here, $\hat{\mathds{1}}_l$ and $\hat{s}_{z,l}$ are the identity operator and 
the $z$-component of the spin operator $\hat{s}_{l}$,
respectively, both acting on electron $l$.
$\delta(\vec{r}_l - \vec{r}\,)$ is the Dirac delta function and $N_{\mathrm{el}}$ is the number of electrons. The plus sign in square brackets of Eq.~(\ref{eq:sigma_dens_oper})
corresponds to the $\hat{\rho}^{\alpha}$ operator, whereas the minus sign produces $\hat{\rho}^{\beta}$.
The resulting density $\rho^{(ij),\sigma}(\vec{r}\,)$ is expressed in terms of non-orthogonal 
MOs $\{\phi_{k}^{(i),\sigma}\}$ and $\{\phi_{l}^{(j),\sigma}\}$ and elements of the transposed 
and inverse overlap matrix $\textbf{S}^{(ij), \sigma}$ such that~\cite{pavanello2011,pavanello_voorhis2013,solo2014,ramos_papa_pavall2015}
\begin{equation} \label{eq:trans-dens2}
\rho^{(ij),\sigma}(\vec{r}\,) = \sum_{k,l=1}^{N_{\mathrm{el}}^{\sigma}} \phi^{(i),\sigma}_k(\vec{r}\,) \left(\textbf{S}^{(ij),\sigma}\right)^{-1}_{lk} \phi^{(j),\sigma}_l(\vec{r}\,). 
\end{equation}
Here, the indices $k$ and $l$ refer to MOs used for construction of Slater determinants $\Phi_i$ and $\Phi_j$, respectively.
Note that these indices are interchanged for the overlap matrix $\textbf{S}^{(ij),\sigma}$ denoting transposition.
Alternatively, this expression can be given as, 
\begin{equation} \label{eq:JTD_frag}
    \rho^{(ij),\sigma}(\vec{r}\,) = 
    \sum_{I,J=1}^{M} \sum_{k \in I } \sum_{l \in J} \phi^{(i),\sigma}_k(\vec{r}\,) \left(\textbf{S}^{(ij),\sigma}\right)^{-1}_{lk} \phi^{(j),\sigma}_l(\vec{r}\,), 
\end{equation}
where the first sum runs over all $M$ subsystems used for the constructions of quasi-diabatic states, while 
the second and third sums run over MOs belonging to subsystems $I$ and $J$, respectively.

After obtaining all necessary components for Eq.~(\ref{eq:eigenvalueproblem}), the linear combination coefficients $B_{im}$ and energies $E_m$ of the adiabatic wave 
functions $\Psi_m$ can be calculated.
The obtained linear combination coefficients $B_{im}$ can be used to construct $n$ adiabatic wave functions $\Psi_m$. From these wave functions other 
molecular properties such as, for example, the spin-density distribution can be calculated~\cite{arti2018,arti2020,arti2021}.
To that end, expectation values of the $\alpha$- and $\beta$-density operators need to be computed,
\begin{equation}
\rho^{\sigma}_m(\vec{r}\,) =\frac{\mel**{\Psi_m}{\hat{\rho}^{\sigma}}{\Psi_m}}{\braket{\Psi_m}{\Psi_m}}
= \frac{\sum_{i,j}^n B_{im}B_{jm}\rho^{(ij),\sigma}S_{ij}}{\sum_{i,j}^n B_{im}B_{jm}S_{ij}}. \label{eq:multi-spin-dens}
\end{equation}
Here, Eqs.~(\ref{eq:linearcombination}), (\ref{eq:overlapelement}), and (\ref{eq:trans-dens2}) were applied. The resulting spin density $\rho^{\alpha - \beta}_m(\vec{r}\,)$
of electronic state $m$ is then calculated as a difference of its $\alpha$- and $\beta$-components,
\begin{equation}
\rho^{\alpha - \beta}_m(\vec{r}\,) = \rho^{\alpha}_m(\vec{r}\,) - \rho^{\beta}_m(\vec{r}\,).
\end{equation}
The spin-density calculated in this way was compared against accurate \textit{ab initio} results~\cite{arti2018} and experimental measurements~\cite{arti2020,arti2021} showing a good qualitative agreement in both cases.

\subsection{Approximate FDE-diab} \label{sec:uncoupled}

As can be seen from Eq.~(\ref{eq:diabaticstate}), the quasi-diabatic states are constructed from Slater determinants of all subsystems' MOs.
Thus, calculations of the inverse overlap matrix $\left(\textbf{S}^{(ij),\sigma}\right)^{-1}$ from Eq.~(\ref{eq:trans-dens2}) and Hamilton matrix elements $H_{ij}$ from 
Eq.~(\ref{eq:hamiltonelement}) are carried out in the orbital space of the total molecular system. The overall computational cost of FDE-diab, therefore, 
grows rapidly with the number of MOs and subsystems considered and becomes prohibitively large for large (bio-)molecular systems. In order to tackle this 
issue, different approximations can be invoked. One of such approaches was proposed in Ref.~\cite{pavanello_voorhis2013} and 
is motivated by large differences between inter- and intra-subsystem MO overlaps.
To demonstrate this, we introduce subsystem MO overlap matrices $\textbf{S}^{(ij), \sigma}_{IJ}$ as (for more detail on the block-structure of $\textbf{S}^{(ij), \sigma}$, 
see Sec.~S1 in the SI),
\begin{equation} 
    \left(\textbf{S}^{(ij), \sigma}_{IJ}\right)_{kl} = \Braket{\phi_{k \in I}^{(i),\sigma}|\phi_{l \in J}^{(j),\sigma}},
\end{equation}
where MOs $\phi_{k \in I}^{(i),\sigma}$ belong to subsystem $I$ and 
correspond to quasi-diabatic state $\Phi_i$. 
Because the elements of intra-subsystem overlap matrices $\textbf{S}^{(ij), \sigma}_{II}$ are typically much larger than those of inter-subsystem counterparts 
$\textbf{S}^{(ij), \sigma}_{IJ}$, the latter can be set to zero giving rise to a modified form of Eq.~(\ref{eq:JTD_frag}),
\begin{equation} \label{eq:pure_DTD}
    \rho^{(ij),\sigma}(\vec{r}\,) \approx \sum_{I=1}^{M} \rho^{(ij),\sigma}_{I}(\vec{r}\,) 
    = \sum_{I=1}^{M} \sum_{k \in I } \sum_{l \in I} \phi^{(i),\sigma}_k(\vec{r}\,) \left(\textbf{S}^{(ij),\sigma}_{II}\right)^{-1}_{lk} \phi^{(j),\sigma}_l(\vec{r}\,). 
\end{equation}
The obtained transition density
$\rho^{(ij),\sigma}(\vec{r}\,)$ is equal to the sum of subsystem 
transition densities $\rho^{(ij),\sigma}_{I}(\vec{r}\,)$, thus, leading to linear scaling (with 
the number of subsystems) of FDE-diab~\cite{pavanello_voorhis2013}. 
However, this approach is not capable of describing inter-subsystem spin-density delocalization and charge-transfer processes. 
This limitation can partially be lifted by  
introducing a subset of subsystems $\mathcal{L}$ (undergoing charge-transfer or featuring spin delocalization) and explicitly considering the contributions from their  
MO overlaps. In other words, we set $\textbf{S}^{(ij), \sigma}_{IJ}$ to zero if both 
requirements $I \neq J$ and $I,J \notin \mathcal{L}$ are satisfied. 
This approach was successfully applied in Ref.~\cite{pavanello_voorhis2013} for charge-transfer simulations occurring between two subsystems.

Alternatively, reduction in the overall computational cost can be achieved by decreasing the number of subsystems $M$, as seen in Eq.~(\ref{eq:JTD_frag}), used 
for the construction of quasi-diabatic states. In this case, the electron density of the total system can still be relaxed within 
FDE by means of freeze-and-thaw cycles~\cite{weso1996}, but
only the MO coefficients of the chosen subsystems are used in the subsequent FDE-diab computation. 
This approximation was previously employed for calculations of spin-density distributions in photosynthetic 
cofactors embedded in large parts of the protein environment and showed good agreement with available experimental results~\cite{arti2020,arti2021}. 
It should also be noted that additional minor computational savings
can be obtained by considering a subset of the constructed quasi-diabatic states in the solution of the generalized eigenvalue problem 
from Eq.~(\ref{eq:eigenvalueproblem}). The above-mentioned techniques can, in principle, be combined producing various approximate FDE-diab computational schemes.
For these calculations, we adopt the FDE-diab($K,L,M$) notation.
Here, $K$ stands for the number of 
quasi-diabatic electronic states considered while solving the generalized eigenvalue problem, $L$ is the number of subsystems 
with non-zero inter-subsystem MO overlaps (from $\mathcal{L}$), and $M$ is the number of subsystems used for 
the construction of quasi-diabatic states. Thus, if a molecular system 
$[ABCDE]\ce{^{.+}}$ is considered, one might solve a 2$\times$2 eigenvalue problem coupling only electronic states  
$\ket{A\ce{^{.+}}BCD}$ and $\ket{AB\ce{^{.+}}CD}$ embedded in environment $E$, where non-zero inter-subsystem MO overlaps are employed for $A$, $B$, and $C$. 
This procedure can then be denoted as 
FDE-diab(2,3,4), while the reference non-approximate FDE-diab computation would correspond to FDE-diab(5,5,5). 
In the following, if not stated otherwise,
non-approximate FDE-diab computations will be denoted simply as FDE-diab.

\section{Computational Details} \label{sec:comput_details}

The structures of the deoxyribonucleic acid (DNA) base triplets guanine--thymine--guanine (GTG) and guanine--adenine--guanine (GAG), molecules from the HAB11 benchmark set, and those presented in Sec.~S2.2 in the SI were taken from 
Refs.~\cite{blancafort2006,kubas2014,arti2018}, respectively, and used without further optimization.
The benzene octamer molecular structure was created with the \textsc{Avogadro} program~\cite{avogadro} and fully optimized using the \textsc{Turbomole} v7.4.1 
program package~\cite{ahlrichs1989} with $C_{2h}$ symmetry. 
In this calculation, the valence polarized triple-$\zeta$ Def2-TZVP basis set~\cite{ahlrichs1992,ahlrichs1994} and 
the XC functional PBE0~\cite{perdew1996,adamo1999} were employed. 
Additionally, the resolution-of-the-identity (RI) approximation in conjunction 
with the auxiliary Coulomb fitting valence triple-$\zeta$ basis Def2-TZV/J~\cite{ahlrichs1992,ahlrichs1994} was enabled. 
Dispersion interactions were taken into account using the D3BJ correction 
with Becke--Johnson damping~\cite{grimme2010_d3,grimme2011_dumping}.

Calculations of spin-density distributions and electronic couplings were carried out 
with \textsc{Serenity}~\cite{serenity} and \textsc{Adf}~\cite{adf_package,jacob2008a} program packages.
All \textsc{Serenity} calculations were performed using the Def2-TZVP  
and auxiliary Coulomb fitting Def2-TZV/J basis sets under the RI approximation.
In the case of \textsc{Adf}, the triple-$\zeta$ TZP~\cite{sto_tzp_basis} monomer basis set from the \textsc{Adf} 
basis set library was used. 
KS-DFT calculations of spin-density distributions were carried out with \textsc{Serenity} using 
the generalized-gradient approximation (GGA) PBE~\cite{pbe}, 
hybrid B3LYP (20\% exact exchange)~\cite{b3,lyp,vwn,b3lyp}, 
range-separated hybrid CAM-B3LYP (19\% exact exchange)~\cite{camb3lyp}, 
and double-hybrid B2PLYP (53\% exact exchange)~\cite{b2plyp} functionals.
In cases of FDE-diab/FDE-ET~\cite{pavanello2011,pavanello_voorhis2013,solo2014,ramos_papa_pavall2015,arti2018} calculations, 
locally modified versions of the \textsc{Serenity}~\cite{serenity} and \textsc{Adf}~\cite{adf_package,jacob2008a} program
packages were applied. In these computations,  
the XC functional PW91~\cite{pw91x_func,pw91} with 
the conjoint~\cite{lee1991} kinetic-energy functional PW91k~\cite{lembarki1994} were adopted.
Relaxation of subsystem densities was accounted for using freeze-and-thaw cycles~\cite{weso1996}. 
Three such cycles were found sufficient to obtain accurate electron densities.
Spin contamination of adiabatic multi-state FDE-diab wave functions was calculated with \textsc{Serenity}
employing the corresponding densities for computations of the $\expval{S^2}$ value (for more details, see Ref.~\cite{wang1995}).

To provide a quantitative measure for the spin-density delocalization, a Becke-type population analysis was carried out with the 
\textsc{Adf} and \textsc{Serenity} programs following the procedure used in our previous works (e.g., see Ref.~\cite{arti2017}). 
To that end, accurate atom-centered Becke grids~\cite{becke1988,becke_grid,becke_grid_adf} were generated. 
Grid points were assigned to particular nuclei forming overlapping atomic-basins. The spin-density values at grid points were integrated over atomic-basins 
yielding atomic spin-density populations. The reported molecular spin-density populations (in percent) were calculated by summing up the 
corresponding atomic contributions over subsystems.

Reference spin-density distributions of 
the $\ce{[GTG]^{.+}}$ and $\ce{[GAG]^{.+}}$ radical cations were computed with CASSCF~\cite{roos1980,ruedenberg1982,olsen2011}. 
These calculations were performed in \textsc{OpenMolcas}~\cite{openmolcas2019,openmolcas2020} 
using Dunning's correlation-consistent 
cc-pVDZ basis set of double-$\zeta$ quality~\cite{dunning1989,woon1993}. Restricted HF
orbitals of neutral molecules around the HOMO--LUMO gaps provided an initial guess for the CASSCF calculations 
of the corresponding radical cations.  
All multiconfiguration calculations were carried out in a state-specific manner. 
The notation CAS($l,k$)-SCF was adopted in this work, where $l$ and $k$ stand for the numbers of active space 
electrons and orbitals, respectively. Converged spin-density distributions were obtained with 
CAS(15,16)-SCF.

\section{Results} \label{sec:results}

In the following, we first present an in-depth comparison of various properties calculated with the two-state FDE-diab implementation in the \textsc{Serenity} program 
and with that from \textsc{Adf} in Sec.~\ref{sec:comp-adf-ser}. Second, the performance of the new multi-state FDE-diab approach for  
spin densities of non-covalently bonded DNA base triplet radical cation complexes is demonstrated 
and compared against KS-DFT and reference CASSCF in Sec.~\ref{sec:dnatriplets}.
Finally, in Sec.~\ref{sec:benzene} we present three approximate multi-state FDE-diab schemes systematically converging to the full FDE-diab results in 
the limit of all couplings, subsystems, and overlaps being considered explicitly.

\subsection{Two-State FDE-diab: Comparison of ADF and Serenity} \label{sec:comp-adf-ser}

To validate the two-state FDE-diab implementation in the \textsc{Serenity} program, 
we calculated a variety of different properties and compared them with those obtained with \textsc{Adf}.
The calculated quantities included: 
i) electronic couplings for 11 dimeric molecular radical cation complexes at different intermolecular displacements 
(benchmarked with \textsc{Adf} in Ref.~\cite{ramos_papa_pavall2015}) 
from the HAB11 set~\cite{kubas2014} and
ii) spin densities and atomic spin populations for dimeric molecular complexes presented in Ref.~\cite{arti2018}.
The complete set of calculated values is given in Sec.~S2 in the SI, 
while here we only briefly discuss the overall two-state FDE-diab performance 
for electronic couplings and atomic spin populations. 
It should be noted that Slater-type basis functions (STOs)
are used in the \textsc{Adf} program, 
whereas Gaussian-type basis functions (GTOs) are employed in \textsc{Serenity}. 
Therefore, small differences between values calculated with \textsc{Adf} and \textsc{Serenity} are expected.

\begin{figure}[!ht]
\centering
\includegraphics[width=\textwidth]{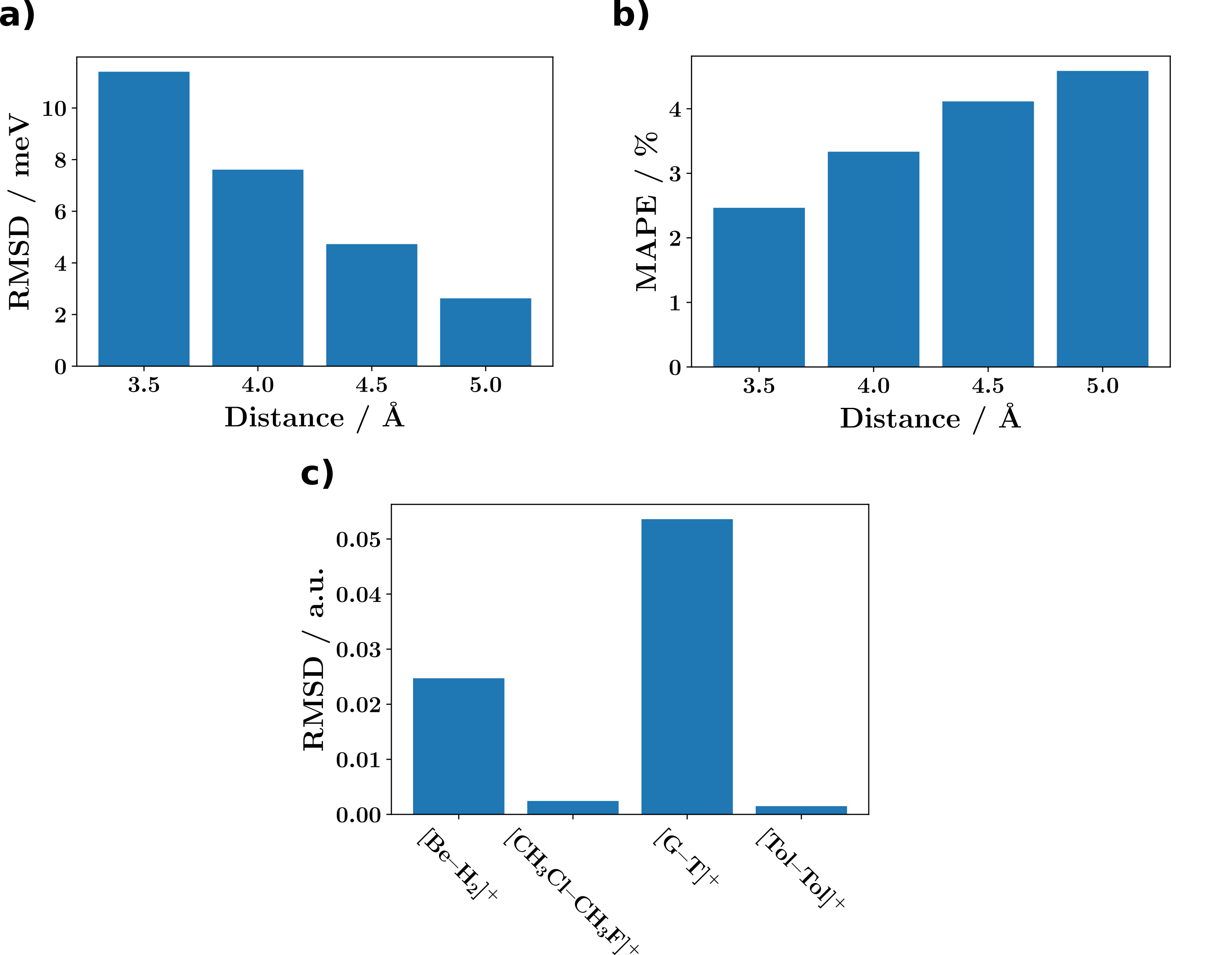}
\caption{Comparison of absolute electronic couplings and atomic spin populations calculated with \textsc{Serenity} and \textsc{Adf}. 
Results shown are a) RMSDs of absolute electronic couplings (in meV) for molecules from the HAB11 benchmark set at different intermolecular distances, 
b) corresponding MAPEs, and 
c) RMSDs of atomic spin populations (in a.u.) for four radical cation complexes from Ref.~\cite{arti2018}.}
\label{fig:validation}
\end{figure}

Root-mean square deviations (RMSDs) of absolute values of electronic couplings calculated with 
\textsc{Serenity} and \textsc{Adf} are shown in 
Fig.~\ref{fig:validation}~a), while mean absolute percentage errors (MAPEs) are presented in Fig.~\ref{fig:validation}~b).
As can be seen from Fig.~\ref{fig:validation}~a), electronic couplings calculated with both program packages are in a good agreement
and show rather small deviations from each other (for values of electronic couplings, see Sec.~S2.1 in the SI). 
Thus, the largest RMSD equal to about $11$~meV is found for the 
intermolecular distance of 3.5~\AA{}. For larger distances, the RMSD decreases and reaches about $3$~meV at the separation of $5.0$~\AA{}.
An opposite trend is observed for MAPE, which increases with the intermolecular distance. However, the largest deviations is equal to only 
about 5\%. 

Atomic spin populations calculated with \textsc{Serenity} and \textsc{Adf} for complexes from Ref.~\cite{arti2018} 
are also in a very good agreement as seen from Fig.~\ref{fig:validation}~c). The largest RMSDs of about 
0.05 and 0.02~a.u.\ are obtained for $\ce{[G-T]^{.+}}$ and $\ce{[Be-H_2]^{.+}}$, respectively. For the other two complexes, deviations
are very small and are below 0.002~a.u. For spin-density isoplots and molecular spin populations, see Fig.~S2 in the SI.
Errors in other quantities calculated with \textsc{Serenity} are also in a good agreement with those from \textsc{Adf} (see Sec.~S2 in the SI). 
Therefore, we conclude the FDE-diab implementation in \textsc{Serenity} established in this work produces very similar results to the reference implementation in \textsc{Adf}~\cite{arti2018,pavanello_voorhis2013} with rather small 
differences caused by different-type basis sets used.

\subsection{Multi-State FDE-diab for DNA Base Triplets} \label{sec:dnatriplets}

Charge- and hole-transfer processes in $\pi$-stacked DNA nucleobases received considerable attention in the literature. 
In the case of small radical cation complexes containing two and three nucleobases, 
many molecular properties such as electronic couplings, charge distributions, and excitation energies are well-studied 
and available (e.g., see Refs.~\cite{blancafort2006,blancafort2007}). 
This makes them very good testing examples for the multi-state FDE-diab framework. 
In the following, we present FDE-diab computations for the radical cation complexes of the DNA 
base triplets $\ce{[G_1TG_2]^{.+}}$ and $\ce{[G_1AG_2]^{.+}}$. 
To distinguish between the two guanine fragments, we label them G$_1$ and G$_2$ in both complexes. 
We compare FDE-diab results against those obtained with KS-DFT and CASSCF.
The CASSCF method has been frequently applied for accurate calculations of spin densities in the past (for 
examples, see Refs.~\cite{arti2017,boguslawski2011}) and serves as the reference in this work. 
In FDE-diab calculations, three quasi-diabatic states of the form $\Phi_1 = \ket{\ce{G_1^{.+}BG_2}}$, $\Phi_2 = \ket{\ce{G_1B^{.+}G_2}}$ 
and $\Phi_3 = \ket{\ce{G_1BG_2^{.+}}}$, where B is adenine or thymine, were constructed. 
Solving the eigenvalue problem in the basis of these quasi-diabatic states, 
three adiabatic wave functions were obtained: The ground state $\Psi_0$ and first two excited states $\Psi_1$ 
and $\Psi_2$. These wave functions were then used to obtain various molecular properties.

\begin{figure}[!ht]
    \centering
    \includegraphics[width=0.75\textwidth]{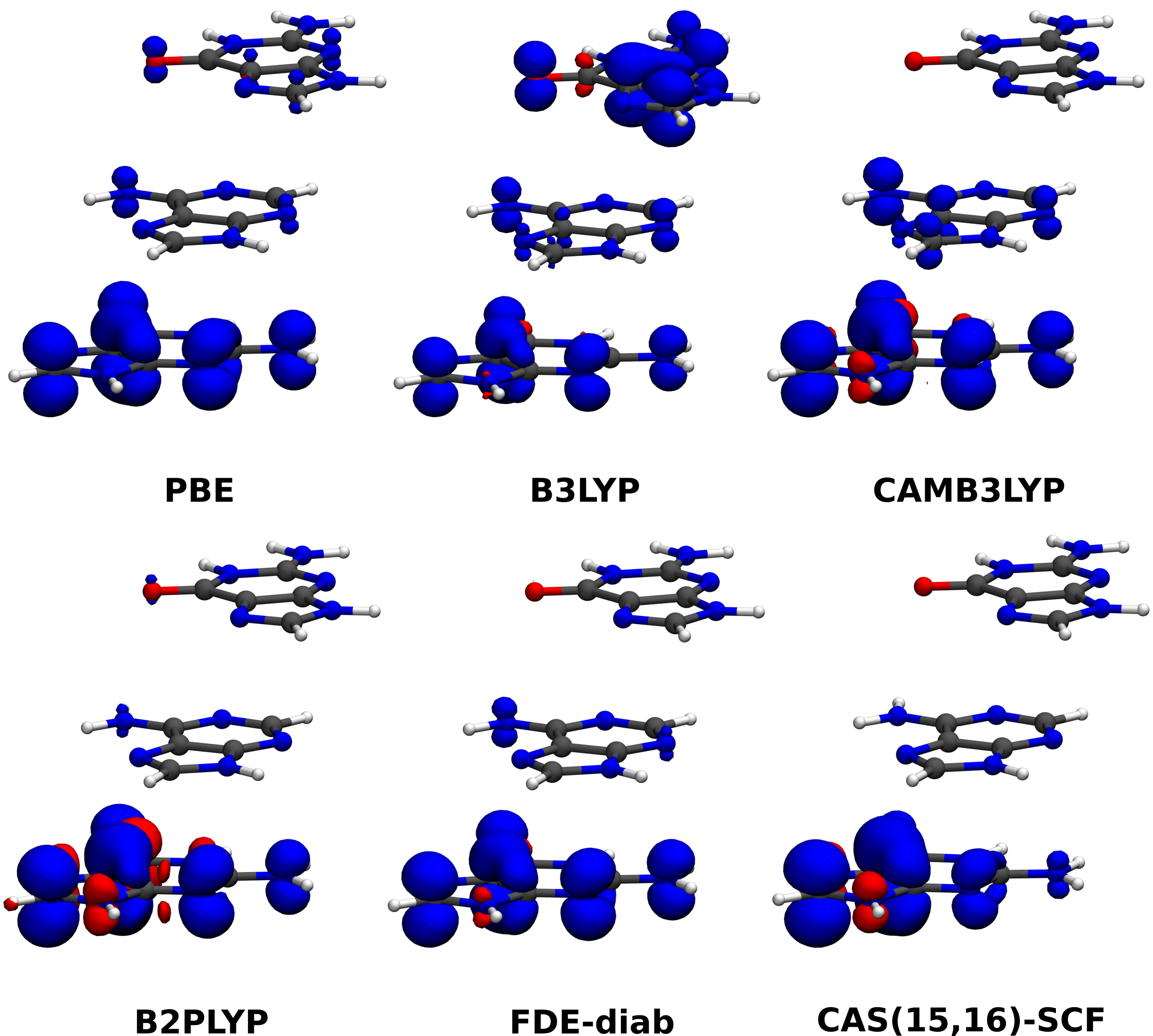}
    \caption{Spin-density distributions (isovalue $\pm 0.003$ a.u.) of the $\ce{[G_1AG_2]^{.+}}$ radical cation complex. 
    Blue and red colors represent positive and negative spin densities, respectively.}
    \label{fig:gag-ksdft}
\end{figure}

\begin{figure}[!ht]
    \centering
    \includegraphics[width=0.75\textwidth]{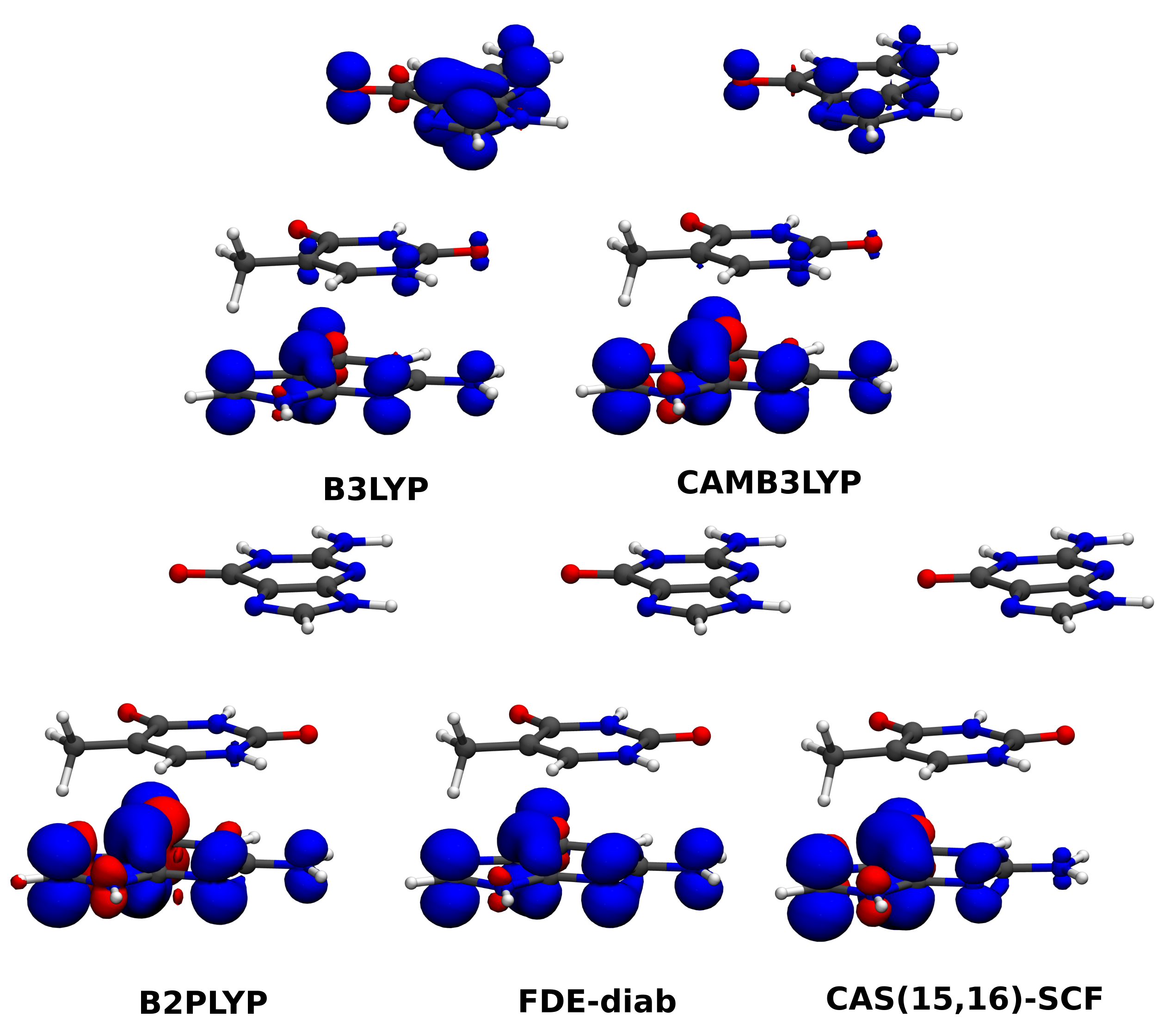}
    \caption{Spin-density distributions (isovalue $\pm 0.003$ a.u.) of the $\ce{[G_1TG_2]^{.+}}$ radical cation complex. 
    Blue and red colors represent positive and negative spin densities, respectively.}
    \label{fig:gtg-ksdft}
\end{figure}

The $\ce{[G_1AG_2]^{.+}}$ and $\ce{[G_1TG_2]^{.+}}$ ground state spin-density distributions 
calculated with KS-DFT, FDE-diab, and CAS(15,16)-SCF are shown in Figs.~\ref{fig:gag-ksdft} and \ref{fig:gtg-ksdft}, respectively, whereas
molecular spin-density populations and $\expval{S^2}$ expectation values are given in Tab.~S15 in the SI.
As one can see, the reference CAS(15,16)-SCF spin densities of $\ce{[G_1AG_2]^{.+}}$ and $\ce{[G_1TG_2]^{.+}}$ are almost fully localized at the \ce{G1} molecules. 
In the both complexes, \ce{A} and \ce{T} carry very small spin-density contributions of about 1--2\%, while 
no spin distribution is found at \ce{G2}. The spin density calculated with KS-DFT is strongly dependent on 
the XC functional used. Similar results were previously reported in several works (for examples, see Refs.~\cite{solo2012,arti2018}).
The double hybrid B2PLYP functional shows the best agreement with the CASSCF reference and produces about 93\% and 95\% of spin 
being localized at \ce{G1} in $\ce{[G_1AG_2]^{.+}}$ and $\ce{[G_1TG_2]^{.+}}$, respectively. However, in this case the corresponding $\expval{S^2}$
expectation values are much larger than the exact value of 0.75~a.u.\ and point out at a high degree of spin contamination. The latter also results in  
overestimated areas of the negative spin density (compared to CASSCF) as seen from Figs.~\ref{fig:gag-ksdft} and \ref{fig:gtg-ksdft}.
This is expected as increasing amounts of the exact exchange often lead to larger spin
contamination, which is directly related to the negative spin density~\cite{herrmann2005,herrmann2006,cohen2007,Radom2008,pavanello2011,arti2017}.
The use of CAM-B3LYP results in somewhat smaller spin contamination values, but strongly overestimates (by about 16--20\%) the spin contributions at    
\ce{A} in $\ce{[G_1AG_2]^{.+}}$ and $\ce{G_2}$ in $\ce{[G_1TG_2]^{.+}}$. The most strongly delocalized spin distributions are calculated with the B3LYP functional.
In the case of PBE, a very good agreement with the reference is found for the $\ce{[G_1AG_2]^{.+}}$ complex.
The corresponding $\expval{S^2}$ value is equal to about 0.85~a.u.\ and only slightly differs from 
the exact value of 0.75~a.u. However, the calculation of $\ce{[G_1TG_2]^{.+}}$ employing the PBE functional does not converge. 
As seen from Figs.~\ref{fig:gag-ksdft} and \ref{fig:gtg-ksdft}, the FDE-diab spin-densities are 
largely localized at $\ce{G_1}$ and, therefore, are in a good agreement with CAS(15,16)-SCF.
Moreover, in the case of $\ce{[G_1TG_2]^{.+}}$ an excellent quantitative agreement (with deviations below 0.3\%) is obtained 
for molecular spin populations (see Tab.~S15). The $\ce{[G_1AG_2]^{.+}}$ molecular spin populations calculated with FDE-diab deviate from 
those computed with CAS(15,16)-SCF by about 3--9\% with the largest error obtained for $\ce{G_1}$. 
For the both molecular complexes, the $\expval{S^2}$ values calculated with FDE-diab are equal to about 0.84~a.u.\ 
and are, therefore, very close to the exact expectation value of 0.75~a.u.
Summarizing the obtained results, we conclude that the new multi-state FDE-diab methodology is more robust than standard KS-DFT 
and, similarly to its two-state predecessor~\cite{arti2018}, leads to qualitatively correct spin distributions.

To show-case the capabilities of multi-state FDE-diab, we also present molecular spin populations and 
total adiabatic energies obtained for the first two electronically excited states of 
$\ce{[G_1AG_2]^{.+}}$ and $\ce{[G_1TG_2]^{.+}}$ in Tab.~\ref{tab:excited_sdens}.
Here, CAS(11,12)-SCF computations reported in Ref.~\cite{blancafort2006} were used as the reference.
Additionally, spin-density distributions for the three lowest electronic states are given in Fig.~S6 in the SI.
As can be seen, in the first excited state of both complexes the FDE-diab spin density is 
mostly localized at $\ce{G_2}$, whereas \ce{T} and \ce{A} feature the largest spin distributions in the second excited state.
These results agree well with CAS(11,12)-SCF computations. An excellent agreement between FDE-diab and CAS(11,12)-SCF molecular
spin populations is obtained in the case of $\ce{[G_1TG_2]^{.+}}$, where the calculated values deviate by up to 
1.5\% and 0.2\% for the first and 
second excited states, respectively. A somewhat worse performance of FDE-diab (compared to CAS(11,12)-SCF) is observed for 
$\ce{[G_1AG_2]^{.+}}$, where deviations of up to about 20\% are found.
In all FDE-diab computations presented, 
the $\expval{S^2}$ value is equal to about 0.81--0.83~a.u. 
Therefore, the calculated spin-density distributions are only slightly spin contaminated.
An opposite trend can be seen for energy values. 
Thus, the first two excited state energies of $\ce{[G_1AG_2]^{.+}}$ calculated with FDE-diab and
CAS(11,12)-SCF deviate by less than 0.01~eV, whereas the second excited electronic state of 
$\ce{[G_1TG_2]^{.+}}$ shows a much larger difference of about 0.5~eV. 
It should be noted, however, that the observed deviations do not necessarily mean a bad performance of FDE-diab as a rather 
small active space and no perturbation energy correction were employed in the reference CASSCF calculations.
On the opposite, qualitative agreements between FDE-diab and CASSCF molecular properties found for the three lowest electronic states is 
rather impressive. This may point out at the possibility to apply the FDE-diab computational approach as a less 
expensive but still accurate enough alternative to very expensive CASSCF for spin-density and energy calculations of large molecular systems.

\begin{table}[!ht]
\centering
\caption{FDE-diab and CAS(11,12)-SCF molecular spin populations $P(\text{nucleobase})$ and total adiabatic energies $E$ (relative to the ground state energy $E_0$) 
of the $\ce{[G_1AG_2]^{.+}}$ and $\ce{[G_1TG_2]^{.+}}$ radical cation complexes in the first two excited 
electronic states ($\Psi_1$ and $\Psi_2$). CAS(11,12)-SCF results are taken from Ref.~\cite{blancafort2006}.}
\begin{tabular}{lcccc} 
\toprule\midrule
Electronic state                   & $P(\ce{G}_1), \%$ & $P(\ce{B}), \%$ & $P(\ce{G}_2), \%$ &  $E$, eV  \\ \midrule
                            \multicolumn{5}{c}{$\ce{[G_1AG_2]^{.+}}$}    \\ \midrule
$\Psi^{\text{FDE-diab}}_1$      &   8.4     &   17.0  &    74.5      & 0.12   \\
$\Psi^{\text{FDE-diab}}_2$      &   3.0     &   75.4  &    21.6      & 0.35   \\
$\Psi^{\text{CASSCF}}_1$        &   1.8     &    4.2  &    94.0      & 0.12    \\
$\Psi^{\text{CASSCF}}_2$        &   3.7     &   90.7  &     5.7      & 0.34     \\ \midrule
                            \multicolumn{5}{c}{$\ce{[G_1TG_2]^{.+}}$}              \\ \midrule
$\Psi^{\text{FDE-diab}}_1$      &   0.2     &    1.5  &    98.2      & 0.21   \\
$\Psi^{\text{FDE-diab}}_2$      &   1.6     &   97.2  &     1.1      & 0.76   \\
$\Psi^{\text{CASSCF}}_1$        &   0.6     &    0.0  &    99.4      & 0.18   \\
$\Psi^{\text{CASSCF}}_2$        &   1.6     &   97.4  &     1.0      & 1.24   \\ \midrule\bottomrule
\end{tabular}
\label{tab:excited_sdens}
\end{table}

Finally, we demonstrate pair-wise electronic couplings calculated with the \textsc{Serenity} program for $\ce{[G_1AG_2]^{.+}}$ and $\ce{[G_1TG_2]^{.+}}$ and compare them against those calculated with the generalized Mulliken--Hush method employing CAS(11,12)-SCF as reported in Ref.~\cite{blancafort2006}. 
It has to be noted here that the FDE-ET module within the multi-state FDE-diab framework in the \textsc{Serenity} program is similar to the original two-state FDE-ET~\cite{pavanello2011,pavanello_voorhis2013,solo2014,ramos_papa_pavall2015} and has no multi-state capabilities. 
To this end, pairs of the generated quasi-diabatic states $\Phi_i$ and 
$\Phi_j$ (described above) were used as the basis for the generalized eigenvalue problem from Eq.~(\ref{eq:eigenvalueproblem}). 
Using the notation system FDE-diab($K, L, M$) introduced in Sec.~\ref{sec:uncoupled}, 
this approach can be denoted as FDE-diab(2,3,3).
The resulting electronic couplings $V_{ij}$ are shown in Tab.~\ref{tab:couplings-comp}.
As one can see, the calculated absolute deviations for the $\ce{[G_1TG_2]^{.+}}$ complex are small 
and vary from about 0.001 to 0.02 eV. Somewhat larger differences of 0.01--0.05~eV are found for $\ce{[G_1AG_2]^{.+}}$.  
As was already stated above, CAS(11,12)-SCF does not employ a large enough active space of orbitals nor an energy 
perturbation correction. Therefore, decisive conclusions on the quality of the presented couplings cannot be made on 
the basis of available results. 
However, at this point we merely aim at demonstrating capabilities of the new extended framework. 
For more thorough benchmarks of electronic couplings, we refer the interested reader to the original works on the 
FDE-ET approach~\cite{pavanello2011,pavanello_voorhis2013,solo2014,ramos_papa_pavall2015}.

\begin{table}[!ht]
\centering
\caption{Absolute values of electronic couplings $|V_{ij}|$ 
for the $\ce{[G_1AG_2]^{.+}}$ and $\ce{[G_1TG_2]^{.+}}$ radical cation complexes. Results are presented for the FDE-diab methodology
used in this work and the generalized Mulliken--Hush method employing CAS(11,12)-SCF from Ref.~\cite{blancafort2006}.
All values are given in eV.}
	\begin{tabular}{lccc}
		\toprule\midrule
		           & FDE-diab  &  CAS(11,12)-SCF  & Difference  \\ \midrule
		  \multicolumn{4}{c}{$\ce{[G_1AG_2]^{.+}}$}    \\ \midrule
		$|V_{12}|$ &  0.063    &   0.049   &  $-$0.014     \\
		$|V_{23}|$ &  0.100    &   0.052   &  $-$0.048      \\
		$|V_{13}|$ &  0.003    &   0.013   &     0.010      \\ \midrule
		  \multicolumn{4}{c}{$\ce{[G_1TG_2]^{.+}}$}    \\ \midrule
		$|V_{12}|$ &  0.090    &   0.078   &  $-$0.012       \\
		$|V_{23}|$ &  0.058    &   0.082   &     0.024       \\
		$|V_{13}|$ &  0.000    &   0.001   &     0.001     \\ \midrule\bottomrule
	\end{tabular}
	\label{tab:couplings-comp}
\end{table}

\subsection{Approximate Multi-State FDE-diab} \label{sec:benzene}

\begin{figure}[!ht]
    \centering
    \includegraphics[width=0.75\paperwidth]{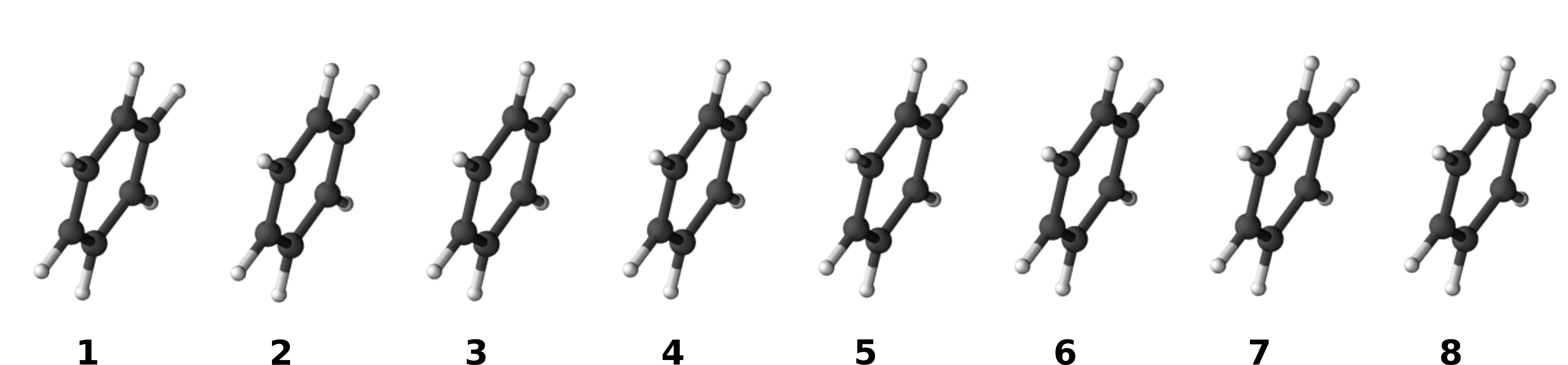}
    \caption{Graphical representation of the $\pi$-stacked benzene octamer. Values denote benzene molecule indices.}
    \label{fig:stack}
\end{figure}

Before demonstrating the efficiency and characteristics of approximate FDE-diab protocols, 
we first introduce our test example of a $\pi$-stacked benzene octamer complex
(see Fig.~\ref{fig:stack}).
For this system, we construct a basis of eight quasi-diabatic states of the form
$\Phi_1 = \ket{\ce{B_1^{.+} B_2 \cdots B_8}}$, $\Phi_2 = \ket{\ce{B_1 B_2^{.+} \cdots B_8}}$, \dots, 
$\Phi_8 = \ket{\ce{B_1 B_2 \cdots B_8^{.+}}}$, where B stands for benzene.
According to the non-approximate FDE-diab approach, i.e., in this case FDE-diab(8,8,8) 
(for the notation, see Sec.~\ref{sec:uncoupled}), 
the largest spin-density contributions of about 29--32\% each are localized on the benzene molecules 
with indices 4 and 5. Smaller molecular spin populations equal to about 14--18\% are obtained for benzene molecules 3 and 6, whereas 
benzene molecules 2 and 7 carry only 3--4\% of the unpaired spin. Negligible amounts of the spin density (below 1\%) are found 
at molecules 1 and 8.

\begin{figure}[!ht]
\centering
\includegraphics[width=0.60\paperwidth]{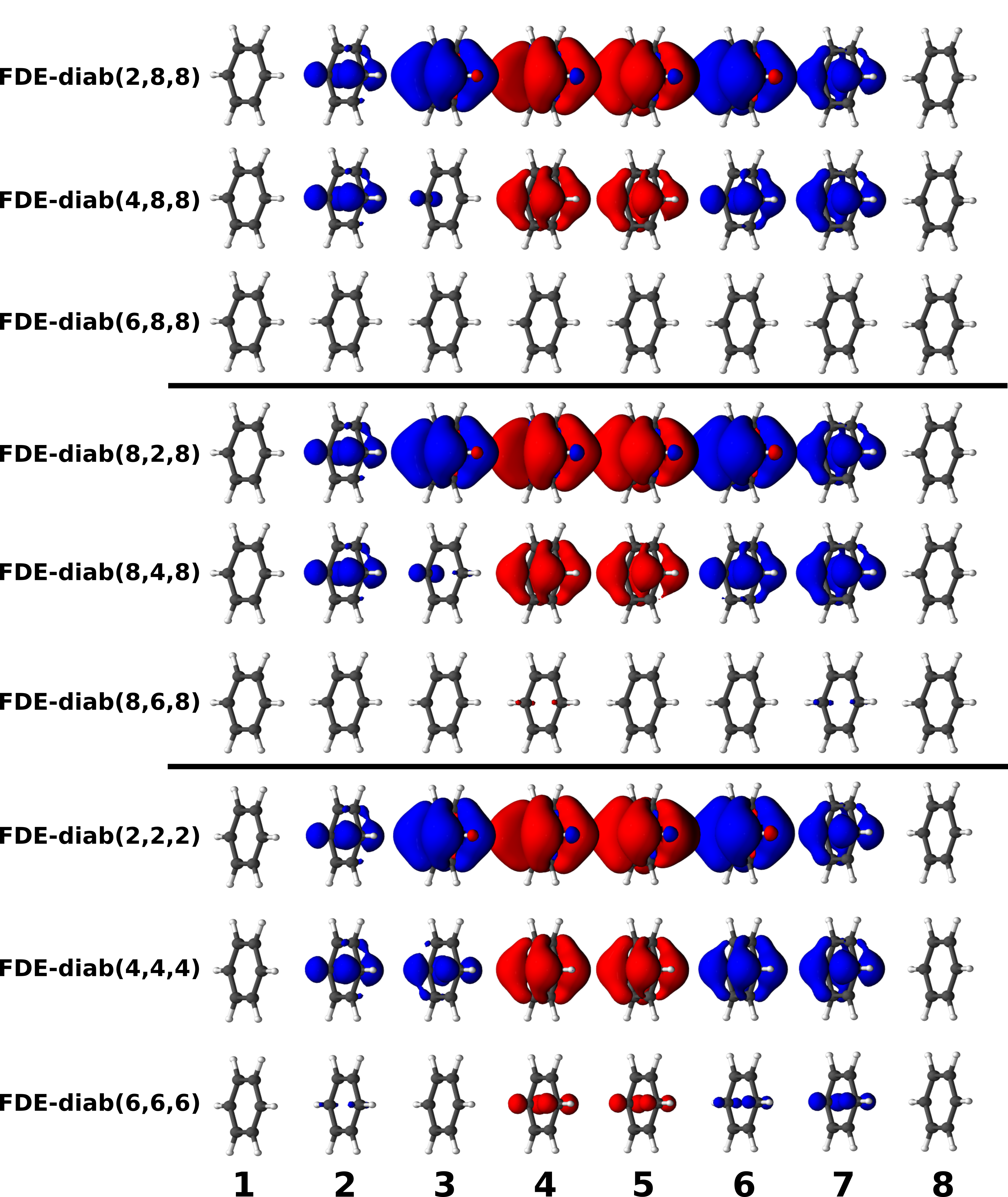}
\caption{Difference ground state spin-density distributions (isovalue: $\pm 0.0003$~a.u.; FDE-diab(8,8,8) spin density 
serves as the reference) calculated for the benzene octamer radical cation using approximate FDE-diab schemes. 
Results are shown for the three series of approximations converging to the full FDE-diab(8,8,8) limit:
(top) FDE-diab($i,8,8$) with $i=2, 4, 6$, (middle) FDE-diab($8,i,8$) with $i=2, 4, 6$, and (bottom) 
FDE-diab($i,i,i$) with $i=2, 4, 6$. Blue and red colors indicate positive and negative difference spin-density 
polarization, respectively. Benzene molecule indices are given with integer numbers.}
\label{fig:approx-combined}
\end{figure}

In principle, many different approximate FDE-diab($K,L,M$) computational protocols can be constructed varying the parameters
$K$, $L$, and $M$ independently. However, here we will only consider those systematically converging to the full 
FDE-diab(8,8,8) solution. To that end, three major strategies can be applied: 
i) varying the number of states $K$ keeping other parameters fixed,
ii) changing the number of subsystems $L$ contained in $\mathcal{L}$ keeping other parameters fixed, and
iii) simultaneously varying all three parameters. 
Note here that the number of subsystems $M$ used for the construction of quasi-diabatic states
defines the maximum values of $K$ and $L$ and, therefore, cannot be changed independently.
Because central benzene molecules 4 and 5 carry the largest spin-density contributions, it is reasonable to start from them
and step-by-step vary the values of $K$, $L$, and $M$ including contributions from the pairs of benzene molecules
which are further away from the center. Keeping this in mind and constructing approximate protocols  
according to the three strategies described above, we introduce three series of approximations:  
FDE-diab($i$,8,8), $i = 2, 4, 6$, 
FDE-diab(8,$i$,8), $i = 2, 4, 6$, and
FDE-diab($i$,$i$,$i$), $i = 2, 4, 6$. 
Using these three series of protocols, we calculate spin-density distributions 
of the benzene octamer and compare them against those computed with FDE-diab(8,8,8).
The difference spin-density distributions are shown in Fig.~\ref{fig:approx-combined}, 
whereas the values of molecular spin populations are given in Tab.~S16 in the SI.

As can be seen from Fig.~\ref{fig:approx-combined} and Tab.~S16 in the SI, 
for $i = 2$ all computational protocols predict about 50\% of the spin density being localized at each of benzene molecules 4 and 5. 
Neighboring molecules 3 and 6 carry only negligibly small amounts of the spin density (below 1\%) for FDE-diab(2,8,8) and FDE-diab(8,2,8), 
while no spin polarization at benzene molecules 3 and 6 is present for FDE-diab(2,2,2). The latter result is not  
surprising and is caused by the fact that only MOs of subsystems 4 and 5 were considered within FDE-diab(2,2,2). 
Already the use of FDE-diab(4,8,8) and FDE-diab(8,4,8) produces spin-density distributions very similar to those from the FDE-diab(8,8,8) computation. 
The corresponding approximate molecular spin populations deviate from the reference ones by less than 1\%. 
Contrary to that, FDE-diab(4,4,4) shows a slower convergence and leads to molecular spin populations deviating from FDE-diab(8,8,8) by up to about 7\%. 
However, this computational scheme produces much smaller deviations (below 2\%) in the next step ($i=6$), while 
the use of other two protocols leads to a slight increase in the error. Overall, all three approaches perform very similar 
and converge to the full FDE-diab solution with increasing $i$.
From the practical point of view, however, FDE-diab($i$,8,8) is of little interest as it leads to the smallest reduction 
in the computational cost compared to the other two schemes.

\section{Conclusions} \label{sec:conclusion}

In this work, we presented a multi-state extension of the recently reported FDE-diab~\cite{arti2018,arti2020,arti2021} approach.
This methodology is based on the FDE formalism and employs non-orthogonal quasi-diabatic states as a basis for 
qualitatively correct calculations of spin densities, electronic couplings, and excitation energies.
We described the new FDE-diab implementation in the \textsc{Serenity} program code and thoroughly tested the method. 
To that end, we first compared the performance of two-state FDE-diab implemented in the \textsc{Serenity} and \textsc{Adf}  
program packages by calculating molecular properties such as electronic couplings, spin densities, and atomic spin populations.
We showed that both implementations produce very similar results with only minor deviations caused by different basis 
set types used (GTO in \textsc{Serenity} vs.\ STO in \textsc{Adf}).

Second, we validated the multi-state FDE-diab framework by calculating ground and excited state molecular properties 
of the DNA base triplets $\ce{[G_1AG_2]^{.+}}$ and $\ce{[G_1TG_2]^{.+}}$ and compared the obtained results against those 
from KS-DFT and the CASSCF reference. As expected, KS-DFT ground state spin-density distributions were strongly dependent on the 
XC functional used. Calculations employing double-hybrid B2PLYP led to the best agreement with CASSCF, but also featured large 
spin contamination values (spin expectation values $\expval{S^2}$ of about 1.1 a.u.) and, as the result, overestimated areas of the negative spin density.
Contrary to that, an impressive agreement with the CASSCF reference was found while using FDE-diab 
with the rather simple GGA-type XC functional PW91. In this case, 
qualitatively correct (compared to CASSCF) spin densites, molecular spin populations, and 
excitation energies were obtained for the three lowest electronic states of $\ce{[G_1AG_2]^{.+}}$ and $\ce{[G_1TG_2]^{.+}}$. 
The corresponding FDE-diab spin expectation values $\expval{S^2}$ were equal to about 0.8 a.u.\ and, therefore, showed only a slight degree of spin contamination. 
Moreover, an excellent quantitative agreement with CASSCF was obtained for 
$\ce{[G_1TG_2]^{.+}}$ molecular spin populations (errors below 1.5\%) and $\ce{[G_1AG_2]^{.+}}$ excitation energies (deviations below 0.01~eV).
It should be noted, however, that a rather small active space and no perturbation energy correction were employed in the reference CASSCF calculations.
Therefore, somewhat larger deviations found for other molecular properties and complexes do not necessarily mean a worse performance of the FDE-diab approach.

Finally, we showed that the new multi-state FDE-diab framework implemented in the \textsc{Serenity} program allows us to 
apply diffent combinations of approximate techniques reported previosly for two-state FDE-diab and FDE-ET (see 
Refs.~\cite{pavanello_voorhis2013} and \cite{arti2020,arti2021}). A unified notation system for these combinations was proposed. Using an example of a 
$\pi$-stacked benzene octamer radical, we demonstrated how these techniques can be applied to construct series of approximate approaches, 
which considerably reduce the overall computational cost and systematically converge to the full multi-state FDE-diab solution.

Summarizing the presented results, we conclude that the new multi-state FDE-diab approach is an effective tool for calculations of molecular properties such as 
spin densities, electronic couplings, and excitation energies. Our test computations demonstrate a surprisingly good agreement with the reference results and hint 
at the potential of FDE-diab to be a less expensive but accurate enough alternative to computationally very demanding CASSCF. 
The approximate techniques implemented 
within the multi-state FDE-diab framework allow one to target large molecular systems of biochemical interest.  
This is due to the favourable scaling behaviour of FDE-diab (and the underlying FDE method) with the number of subsystems~\cite{serenity}.
Two-state FDE-diab calculations of photosynthetic reaction center models including large parts of the protein environments can serve as examples
for large-scale calculations with this technique~\cite{arti2020,arti2021}.
However, the developed methodology is based on FDE and, therefore, shares the same set of limitations including 
the inability to describe convalently bonded or strongly interacting fragments.
This limitation is lifted by methods such as potential reconstruction~\cite{leeuwen1994,wu2003,zhang2018}, projection-based embedding~\cite{huzinaga1971,manby2012,khait2012} and three-partition FDE~\cite{jacob2008,fux2010,kiewisch2013}. Whether these techniques can effectively be combined with the FDE-diab methodology is an open question, which will be addressed elsewhere.

\section*{Acknowledgments}
P.E. thanks Niklas Niemeyer for discussions on the \textsc{Serenity} framework 
and Dr.\ Milica Feldt for her help with CASSCF calculations.
D.G.A. acknowledges funding from the European Union's Horizon 2020 research and innovation
programme under the Marie Sk\l{}odowska--Curie grant agreement no.\ 835776.

\newpage
\clearpage

\end{document}